\documentclass[12pt]{article}
\usepackage{amsmath, amssymb, graphics}

\usepackage{lineno}

\usepackage{graphics, psfrag}
\usepackage{graphicx}
\usepackage{mathrsfs}

\def\id{\protect{{1 \kern-.28em {\rm l}}}}

\def\be{\begin{eqnarray}}
\def\ee{\end{eqnarray}}

\makeatletter
\renewcommand\section{\@startsection {section}{1}{\z@}%
								   {-3.5ex \@plus -1ex \@minus -.2ex}%
								   {2.3ex \@plus.2ex}%
								   {\normalfont\large\bfseries}}
\renewcommand\subsection{\@startsection{subsection}{2}{\z@}%
								   {-3.25ex\@plus -1ex \@minus -.2ex}%
								   {1.5ex \@plus .2ex}%
								   {\normalfont\normalsize\bfseries}}
\usepackage{newfloat}
\DeclareFloatingEnvironment{Supplementary Figure}

\makeatother

\newcommand{\mathsym}[1]{{}}
\def\k{\kappa}

\let\a=\alpha \let\b=\beta \let\g=\gamma \let\d=\delta 
\let\z=\zeta  \let\th=\theta  \let\k=\kappa
\let\l=\lambda \let\m=\mu \let\n=\nu \let\x=\xi  
 \let\t=\tau    
 \let\vep=\varepsilon
	    \let\D=\Delta \let\Th=\Theta \let\L=\Lambda
  \let\S=\Sigma  
 
\let \x = \chi
\let \z = \xi

\let\la=\label \let\ci=\cite 
  
 \def\bd{\begin{document}} \def\ed{\end{document}}
\def\ds{\documentstyle} \let\fr=\frac \let\bl=\bigl \let\br=\bigr

\def\td{\tilde}
\newcommand{\ra}{\rightarrow}
\newcommand{\lra}{\longrightarrow}
\newcommand{\Lra}{\Leftrightarrow}
\newcommand{\Tr}{{\rm Tr} }

\def\ve{\varepsilon}
\def\vf{\varphi}
\def\wg{\wedge}

\def\rmp{Rev. Mod. Phys.}
\def\pasa{Pub. Astron. Soc. Australia}
\def\cqg{Class. Quantum Grav.}
\def\aap{Astron. Astrophys.}
\def\aaps{Astron. Astrophys. Suppl. Ser.}
\def\mnras{Mon. Not. R. Astron. Soc.}
\def\apjs{Astrophys. J. Suppl. Ser.}
\def\apjl{Astrophys. J. Lett.}
\def\apj{Astrophys. J.}
\def\nat{Nature}
\def\araa{Ann. Rev. Astron. Astrophys.}
\def\aj{Astronom. J.}
\def\physrep{Phys. Rep.}
\def\prd{Phys. Rev. D}
\def\prc{Phys. Rev. C}
\def\prl{Phys. Rev. Lett.}
\def\nphysa{Nuc. Phys. A}
\def\jqsrt{J. Quant. Spectrosc. Radiat. Transfer}

\newcommand{\wt}{\widetilde}
\def \foot {\footnote}
\def \bi{\bibitem}

\def \tr {{\rm tr}}
\def \td {\tilde}
\def \ci{\cite}
\def \N {{\mathcal N}}
\def \const {{\rm const}}
\def \t {\tau}
\def\S{{\mathcal S} }

\def \lra {\leftrightarrow}

\def \d {\del}
\def \L {\Lambda}
\def\z{\zeta}

\def \del{\partial}

\def\b{\beta}
\def\l{\lambda}
\def\eps{\epsilon}
\def\vep{\varepsilon}

\def \k {\kappa}

\def\th{\theta}
\def\Th{\Theta}
\def\vth{\vartheta}

\def\vth{\vartheta}
\def\ra{\rightarrow}

\def\ni{\noindent}

\def \la {\label}
\def \eps {\epsilon}

\def \k {\kappa}
\def \t {\tau}

\def \Msun {M_{\odot}}
\def \erg {{\rm erg}}
\def \max {\mathrm{max}}
\def \min {\mathrm{min}}
\def \EM {\mathrm{EM}}
\def \Brg {\mathrm{Br \gamma}}
\def \Ha {\mathrm{H 30 \alpha}}
\def \yr {\mathrm{yr}}
\def \arcsec {\mathrm{arcsec}}
\def \pc {\mathrm{pc}}
\def \kms {\mathrm{km \, s^{-1}}}
\def \cm {\mathrm{cm}}
\def \M  {\mathscr{M}}
\def \K {\mathrm{K}}
\def \vol {\mathrm{vol}}
\def \Sch {\mathrm{s}}
\def \sec {\mathrm{s}}

\begin{document}


\overfullrule=0pt
\parskip=2pt
\parindent=12pt
\headheight=0in \headsep=0in \topmargin=0in \oddsidemargin=0in

\vspace{ -3cm}
\thispagestyle{empty}
\vspace{-1cm}

\begin{center}
\vspace{1cm}
{\Large\bf
A Cool Accretion Disk around the Galactic Centre Black Hole
}

\vspace{.2cm}

Elena M. Murchikova$^{1,2}$,
E. Sterl Phinney$^{2}$,
Anna Pancoast$^{3}$,
Roger D. Blandford$^{4}$


{\footnotesize $^{1}$Institute for Advanced Study, Einstein Drive, Princeton, NJ 08540, USA\\
$^{2}$Theoretical Astrophysics, California Institute of Technology, MC 350-17, Pasadena, CA 91125, USA\\
$^{3}$Harvard-Smithsonian Center for Astrophysics, 60 Garden Street, Cambridge, MA 02138, USA\\
$^{4}$Kavli Institute for Particle Astrophysics and Cosmology, Stanford University, Stanford, CA 94309, USA\\
}



\end{center}

\setcounter{equation}{0}
\setcounter{footnote}{0}
\setcounter{section}{0}





{\bf

A supermassive black hole Sagittarius A* (SgrA*) with the mass $M_\mathrm{SgrA*} \simeq 4 \times 10^6 M_{\odot}$
resides at the centre of our galaxy {\cite{Boehle16,Gravity18}}. 
A large reservoir of hot ($10^7 \, \K$) 
and cooler ($10^2 - 10^4 \, \K$) gas surrounds it within few pc \cite{Genzel10}.
Building up such a massive black hole within the $\sim 10^{10}$ year lifetime of our galaxy would require a mean accretion rate of $\sim 4 \times 10^{-4} \, \Msun \, {\yr}^{-1}.$ 
At present, X-ray observations constrain the rate of hot gas accretion at the Bondi radius ($10^5 \, R_\mathrm{Sch} = 0.04 \, \pc$ at 8 kpc)  to
 $\dot{M}_\mathrm{Bondi} \sim 3 \times 10^{-6} M_{\odot}\,\mathrm{yr}^{-1}$ \cite{Baganoff03,Quataert02,Quataert04}, and 
 polarization measurements \cite{Bower03} constrain it near the event horizon to $\dot{M}_\mathrm{horizon} \sim 10^{-8} M_{\odot} \, \mathrm{yr}^{-1}.$
A range of models was developed to describe the accretion gas onto an underfed black hole {\cite{Narayan.Yi95,Blandford.Begelman99,Quataert.Gruzinov00}}.
However, the exact physics still remains to be understood. 
One challenge with the radiation inefficient  accretion flows (RIAFs) is that even if one understands the dynamics there is no accepted prescription 
for associating emissivity (and absorption) with the flow.
The other issue is the lack of model-independent probes of accretion flow at intermediate radii (between few and $\sim 10^5 R_\Sch$),  i.e. the constraints that do not assume a model of accretion flow as an input parameter.
Here we report a detection and imaging of the $10^4 \, \K$ ionized gas disk within $2 \times 10^4 \, R_\mathrm{Sch}$  in
a millimetre hydrogen recombination line H30$\alpha:$ $\mathrm{n} = 31 \to 30$ at 231.9 GHz {\cite{LR71,Scoville.Murchikova13}}
using the Atacama Large Millimeter/submillimeter Array (ALMA). The emission was detected with
a double-peaked line profile spanning full width of $2,200 \, \kms$ with the approaching and the receding components straddling Sgr~A*, each offset from it by $0.11 \, \arcsec= 0.004 \, \mathrm{pc}.$
The red-shifted side is displaced to the north-east, while the blue-shifted side is displaced to the south-west.
The limit on the total mass of ionized gas estimated from the emission is  $10^{-4} - 10^{-5} \, \Msun$ at a mean hydrogen density $10^5-10^6 \, \cm^{-3},$
depending upon whether or not we assume the presence of a uniform density disk or an ensemble of orbiting clouds, and the amplification factor of the mm radiation due to the strong background source which is Sgr~A* continuum.
}

\bigskip

\begin{figure}[p]
\centering
\includegraphics[width=15cm]{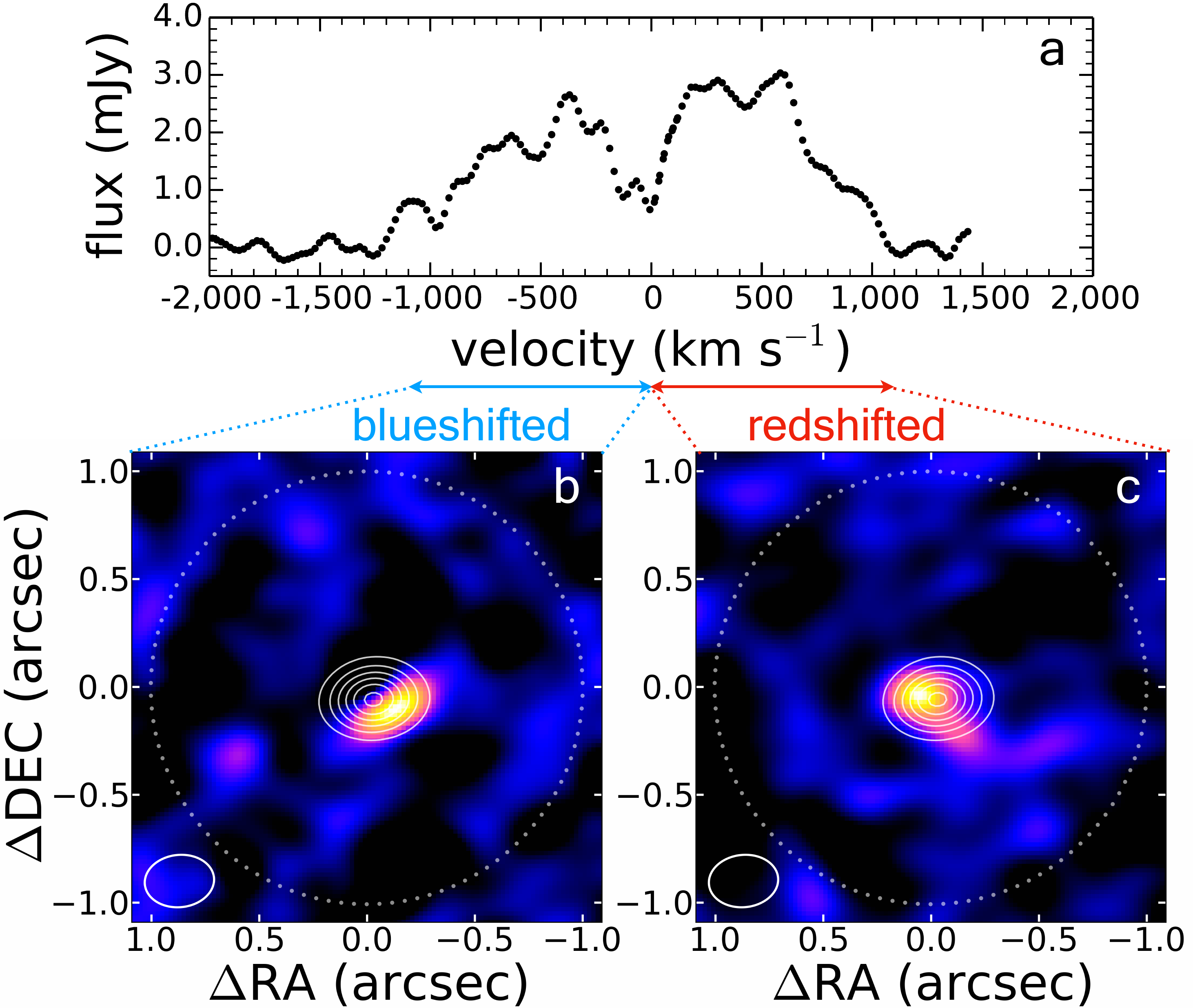}
\caption{ \textbf{The ionized gas emission within $20,000 \, R_s$ around Sgr~A*.}
{\textbf{(a)}} The spectrum of $\mathrm{H}30\alpha$ integrated over 0.23 arcsec (0.009 pc) radius around Sgr~A* detected with ALMA Cycle 3. The rest frequency of $\mathrm{H}30\alpha$ is at zero velocity. Points are averaged over $45\, \mathrm{km \, s^{-1}}.$
The observational uncertainties are $\pm$0.3 mJy, more detail in the Supplementary Information.
{\textbf{(b,c)}} The spatial distribution of the blueshifted and redshifted $\mathrm{H}30\alpha$ emission. The white contours (0.5, 1.0, 1.5, 2.0, 2.5, 3.0, and 3.5 Jy) indicate the continuum emission from SgrA* at $\sim 230$ GHz. The white dotted circle indicates the Bondi radius. $\D$RA and $\D$DEC are the offsets in arcsec from the position of SgrA*. }
\label{spec3v}
\end{figure}

Black hole accretion and feedback are crucial to understanding the evolution of galaxies, the origin of relativistic jets, and to study physics near black hole horizons. 
Sgr~A* is our nearest supermassive black hole. 
As such it offers an excellent opportunity to study accretion and outflow processes close to a black hole.
X-ray measurements of gas density and temperature at the outer edge of the accretion flow, together with the assumptions of spherical adiabatic and constant in time accretion \cite{Bondi52}, 
can be used to estimate the mass supply to the black hole as 
$\dot{M}_\mathrm{Bondi} \sim 3 \times 10^{-6} M_{\odot}\, {\textrm{yr}}^{-1}$ \cite{Baganoff03,Quataert02,Quataert04}.
If the radiative efficiency were $\sim 10\%$ \cite{Thorne74,Novikov.Thorne73}, the luminosity would be roughly 30,000 times the bolometric luminosity of 
$\sim 3 \times 10^{36} \, {\textrm{erg} \, \textrm{s}^{-1}}$ {\cite{Mahadevan98}}.
Extensive theoretical efforts aimed at resolving this discrepancy {\cite{Narayan.Yi95,Blandford.Begelman99,Quataert.Gruzinov00}} have concluded that the accretion proceeds via radiatively inefficient accretion flow (RIAF) in a geometrically thick torus \cite{Phinney81,BBRP82}, in which the electrons remain cooler than the protons,
allowing the observed luminosity to be produced by a much higher accretion rate than in a thin disk (see Supplementary Material), 
while also explaining the mm to $\gamma-$ray spectrum \cite{Narayan.Yi.Mahadevan95}.
However it is difficult to favor or rule out any of the RIAF models for Sgr~A*.
One problem with the RIAFs is that even if one understands the dynamics there is no accepted prescription 
for associating emissivity (and absorption) with the flow. The other is the lack of model-independent observational constraints, i.e. the constraints that do not assume a model of accretion flow as as input parameter, on the gas behavior between $\sim 10 - 10^5 \, R_\mathrm{Sch}.$ \footnote{Constraints on the average density of the accretion flow from the drag on the G2-object's orbit \cite{Gillessen18} employs the density scaling with radius of the accretion flow as $\sim r^{-1}$ and assumes the flow is not rotating, and thus it is also model-dependent.}

To date, our main source of information about the structure of the accretion flow around  Sgr~A* as a whole has been X-ray observations by Chandra, which probe gas at $T \geq10^7$~K and have a resolution of $\sim 1$ arcsec, corresponding to $10^5 \, R_\mathrm{Sch} = 0.04 \, \pc$ at the Galactic Centre \cite{Baganoff03}. This hot gas could, if clumpy, cool to $T\ll 10^7 \, \K,$  becoming invisible in X-rays, while feeding the black hole. There is also a large reservoir of $T\ll 10^7 \, \K$ gas near the Galactic Centre. The circumnuclear disk contains molecular gas at $\sim 10^2 - 10^3$ K and the streamers of the mini-spiral contain $10^4$ K ionized gas near the black hole and cooler $10^2$ K gas on the outside \cite{Genzel10}. Both could supply gas to the black hole. 
This $T\ll 10^7 \, \K$ gas is invisible in X-rays and its contribution to the accretion flow onto Sgr~A* has previously been unconstrained. This gas can be seen in recombination lines of hydrogen \cite{Scoville.Murchikova13}. 

We recently conducted high precision observations with Atacama Large Millimeter/submillimeter Array (ALMA) of the Galactic Centre at 1.3 mm (231.9 GHz) to measure H30$\alpha,$ a recombination line of hydrogen produced in the transition $\mathrm{n}=31 \to 30$ (see  Supplementary Information for detail of the observations). This line was selected because it is free from strong molecular emission  \cite{splatalogue16}.

We detected a double-peaked line centred at the frequency of H30$\a,$ with a full velocity width of $2,200 \, \kms$ (Fig. 1a). The spectrum is integrated within an aperture of $0.63 \times 0.51 \, \arcsec^2.$ The spatial distribution of the redshifted and blueshifted sides of H30$\a$ emission are shown in Fig.~1bc.  Most of the emission is contained within a radius of $0.23 \, \arcsec$ from Sgr~A*, whose continuum emission is indicated with white contours. 
The spatial and spectral structure of H30$\a$ is consistent with a rotating disk (or, less probably, a bipolar outflow; 
for modelling see Supplementary Information). The position angle of the line of nodes is 64 degrees north through east. The absence of  emission in the centre indicates a hole in the disk with a radius similar to the beam size  ($\sim 0.3 \times 0.2 \, \arcsec^2$ FWHM).
The maxima of the blue and red emissions are at $0.11$ arcsec from Sgr~A*.
Assuming a black hole mass of $4 \times 10^6 M_{\odot},$ the Keplerian velocity around Sgr~A* at this radius is $2,000 \, \kms,$  while
the peak of the spectrum is at $\sim 500 \, \kms.$ This implies that the disk is either close to face-on (inclined at an angle $\sim15$ degrees) as was suggested to be the configuration at a few Schwarzschild radii ($R_\Sch$) \cite{GRAVITY18}, or is rotating more slowly than the Keplerian velocity, which would be possible for 
an ensemble of magnetically- or pressure-supported clouds 
in a geometrically thick accretion torus, where the $T\ll 10^7 \, \K$ cloudlets are embedded in a hot accretion flow or magnetocentrifugal wind \cite{Phinney81,BBRP82,Narayan.Yi95,Blandford.Begelman99,Quataert.Gruzinov00}.
Note, that rotation axis of the accretion flow can vary with the radius for many reasons  \cite{Ressler18}. The rotation of the flow at $\sim 10^4 R_{\Sch},$ which is reported here, may differ from the orientation of the flow at a few $R_\Sch$ suggested by GRAVITY observations \cite{GRAVITY18}.

The plane of the $10^4$ K disk  does not seem to coincide with the planes of the other structures in the Galactic Centre (Fig.~2). However, its direction of rotation is similar to that of the Galaxy, the circumnuclear disk, the mini-spiral and the  clockwise stellar system, so the disk could be fed by any of these structures (e.g. \cite{Goicoechea18}). On a smaller scale, the S2 star rotates in the opposite direction to the disk and seems to pass through its plane nearly orthogonally in a third plane. 
The orbit of the G2 object is also inclined to the disk at about 90
degrees, and they are not in similar planes. The orbit of G1 may be roughly coplanar with the disk \cite{Witzel17,Witzel14,Plewa17}. 
 
\begin{figure}[p]
\centering
\includegraphics[width=15cm]{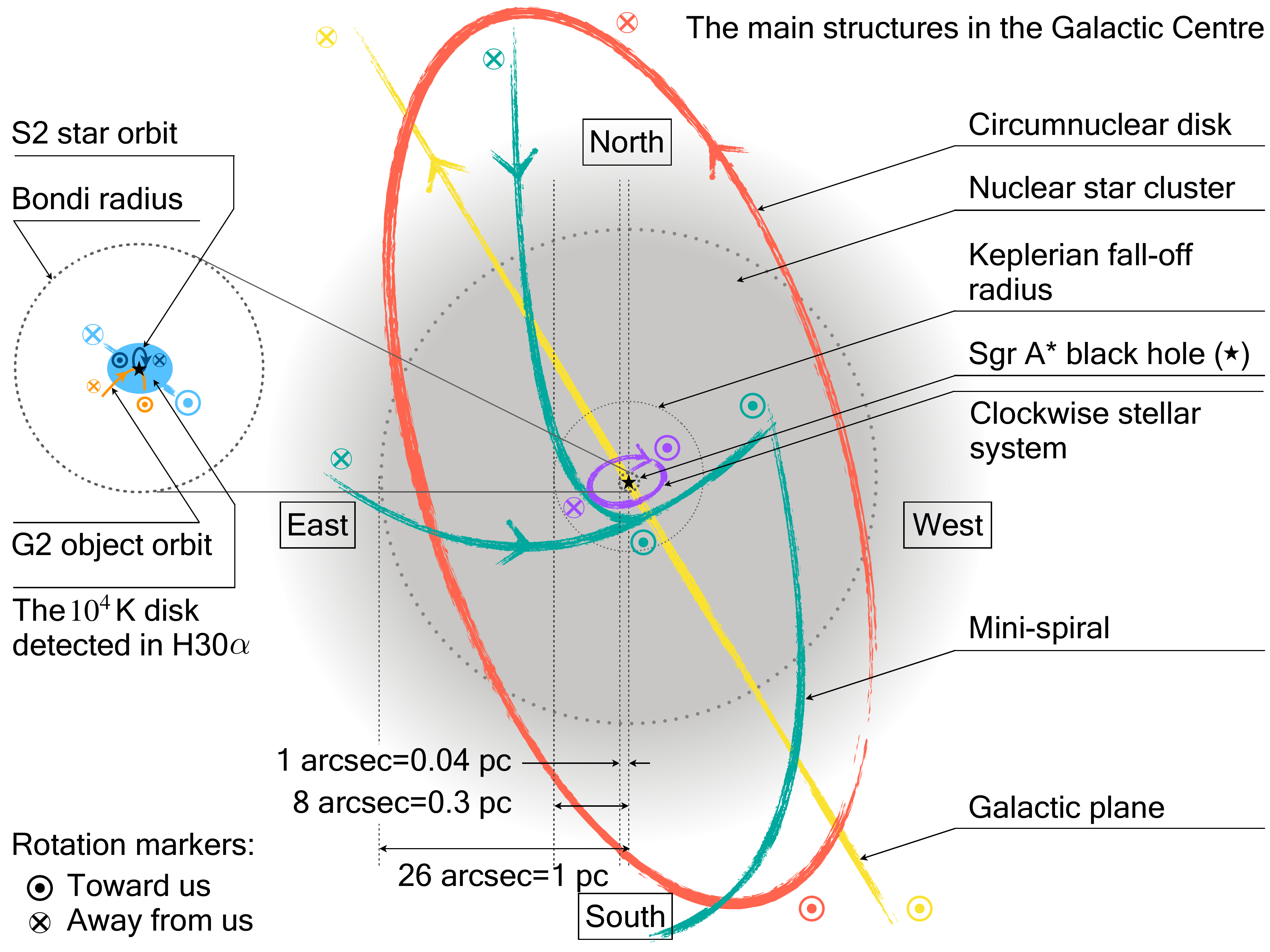}
\caption{
\textbf{The inner two parsec of the Galactic Centre region.}
Schematic to scale plot. The main structures are indicated by different colors. The projected distances in parsec (pc) were calculated for a distance $D_\mathrm{SgrA*}=8.0$ kpc {\cite{Boehle16,Gravity18}}.
\textbf{Black star}: The Galactic Centre black hole Sagittarius A*. 
\textbf{Red oval}: The molecular ring, or circumnuclear disk (torus), containing molecular gas. It is a circle of radius $\sim2$ pc inclined to the line of sight.  
\textbf{Cyan structure}: The mini-spiral consists of three streams of ionized gas rotating counterclockwise orbiting the black hole and/or inflowing \cite{Zhao16}. 
\textbf{Filled grey circle}: The nuclear star cluster. A mass $\sim 1.0 \times 10^6 M_{\odot}$ is concentrated within a radius of about 1 pc. The outer boundary extends to a few pc. 
\textbf{Yellow line}: The edge of the galactic plane.  
\textbf{The Keplerian fall-off radius} 0.3 pc is the radius at which the point mass of Sgr A* starts visibly dominate the velocity dispersion of the stars in the surrounding cluster \cite{Schodel09}.
\textbf{The Bondi radius} for X-ray emitting gas at $10^{7} \, \K$~is $\sim 0.04 \, \textrm{pc}$ \cite{Baganoff03}.  The structures within the Bondi radius are plotted in the zoomed-in region on the left.
\textbf{Dark blue oval}: The orbit of star S2 as projected on the sky.
\textbf{Orange line}: The observed track of G2 object as projected on the sky.
\textbf{Filled light blue oval}~in the centre right around Sgr~A*: The ionized gas disk  detected in H30$\a$ line. The direction of rotation of the structures, where relevant, are shown with arrows, the circled cross indicates recession from Earth, and the circled dot indicates approach. The colors of the rotation markers coincide with the colors of the rotating structures.
The projected distances in parsec (pc) were calculated for a distance to the Galactic Centre $D_\mathrm{SgrA*}=8.0$ kpc {\cite{Boehle16,Gravity18}}. At this distance 1 arcsec corresponds to $0.04$ pc.
}
\label{schematic}
\end{figure}

The velocity-integrated line flux of H30$\alpha$ $S \Delta V_{H30\alpha}$ (the integral under the line in Fig.~1a) allows us to constrain the physical properties of the line-emitting gas
 (Supplementary Information)
\be\label{storey}
    S\D V_{\textrm{H}30\a}  =  \frac{ \epsilon_{\textrm{H}30\a} (T,n)}{4 \pi D^2}  \EM_{\mathrm H30\a} \frac{c}{\nu_{\textrm{obs}}},
\ee
where $\EM_{\mathrm H30\a}= \M \EM$ is effective volume emission measure, $\M$ is magnification factor described below, $\EM = \int n_e n_p d\mathrm{vol} \sim n^2 \times \vol$ is the [true] volume emission measure with $\vol$ being the volume of space occupied by the ionized gas, $\nu_{\textrm{obs}}=231.9 \, \mathrm{GHz}$ is the frequency of the $\Ha$ line,
$D = 8 \, \mathrm{kpc}$ is the distance from the emitting source, and $c$ is the speed of light.
The emissivity of H$30\alpha,$ $\epsilon_{\textrm{H}30\a} (T,n),$ is tabulated for uniform media of constant density with no background radiation \cite{Storey.Hummer95}.
In the case discussed here, however, the effects on the optical thickness ($\tau$) and a factor for maser amplification 
due to the central Sgr~A* background source ($\mu$) must be important and the true volume emission measure ${\EM }$ would be different from 
the effective one $\EM_\Ha$ obtained from Equation \ref{storey}. (Tabulated values \cite{Storey.Hummer95} include a non-equilibrium level population, but no pumping by a background source.)
To account for this, we introduce a parameter $\M\sim \mu (1+\tau)^{-1}$ as the proportionality coefficient between the effective $\EM_{\mathrm H30\a}$ and the true $\EM$ volume emission measure.
The gas cools very efficiently down to $10^4$ K, below which cooling substantially slows. 
The temperature of the gas in the disk is therefore expected to be $\sim 10^4$ K. The tabulated emissivity depends weakly on the density. Since there is only a 30\% change
in $\epsilon_{\textrm{H}30\a}$ over the density range $n=10^4 - 10^7 \, \cm^{-3}$ \cite{Storey.Hummer95}, we will use $\epsilon_{\textrm{H}30\a} (10^4 \,\K, 10^{6} \, \cm^{-3}) = 1.25 \times 10^{-31} 
\, \mathrm{erg \, s^{-1} \, cm^3}$ as a fiducial value (Supplementary Information). 

From $S \Delta V_{H30\alpha}=3.8 \, \textrm{Jy} \, \textrm{km} \, \textrm{s}^{-1},$ where $1 \, \textrm{Jy} = 10^{-23} \frac{ \textrm{erg} }{ \textrm{s}\, \textrm{cm}^{2} \, \textrm{Hz}},$ we obtain the effective volume emission measure of the ionized hydrogen in the disk 
\be\label{em}
   \EM_{\mathrm H30\a} = \M {\EM}  = 1.8 \times 10^{60} \, \textrm{cm}^{-3}.
\ee

We constrain the value of $\M$ using the absence of the double-peaked profile in $\Brg$ in the Galactic Centre observations by A. Ciurlo et al (in prep.), which sets an approximate limit $S_\Brg < 0.1$ mJy from within $0.3 \, \arcsec $ of Sgr~A*, covering our H30$\a$ region. S. Gillessen obtained a similar limit (private communication).
$\Brg$ originates from lower energy levels $\mathrm{n}=7\to4$ with much higher transition rates than H30$\a.$ It should be populated according to Case B recombination and the tabulated emissivity values \cite{Storey.Hummer95} should be valid: $\EM_\Brg \simeq \EM.$
\be
	\M \simeq \frac{\EM_{\mathrm H30\a} }{\EM_\Brg} = \frac{\epsilon_\Brg}{\epsilon_\Ha} \frac{\nu_\Ha}{\nu_\Brg} 10^{-\frac{A_K}{2.5}} 
	\frac{\langle S_{\mathrm{H}30\a}\rangle}{\langle S_{\mathrm{Br \gamma}}\rangle} \geq {4.5} \times \frac{1.7 \, \mathrm{mJy}}{0.1 \, \mathrm{mJy}} \simeq 80.
\ee
Here $A_{K}=2.46$ is the extinction toward the Galactic Centre at the frequency of $\Brg$ \cite{Witzel18},  $\epsilon_\Brg (10^4 \,\K, 10^{6} \, \cm^{-3})=3.33 \times 10^{-27} \,  \mathrm{erg \, s^{-1} \, cm^3}$, $\nu_\Brg=1.39 \times 10^5$ GHz, and  $\langle S_{\mathrm{H}30\a}\rangle =1.7$~mJy is the mean value of $\Ha$ flux across the line.
The relation above implies that H30$\a$ has to be amplified by a factor $\M \gtrsim 80.$ 
Such a large amplification is possible. 
The $10^4 \, \K$ disk is likely heated by magneto-viscous dissipation and external radiation, most relevantly the mm continuum from Sgr~A*. The continuum has a brightness temperature of $\sim 10^4$ K and a spectrum peaking near the frequency of our observed line, and thus may create population inversions.
At $\mathrm{n} \sim 30$ and at the disk densities inferred below, collisions, spontaneous and stimulated emissions, and absorption are all important and can all enhance the H30$\alpha$ emissivity. The expected ionizing flux deduced from ionization equilibrium is consistent with the presence of amplification (Supplementary Information). An exact calculation of the relative fluxes requires further investigation.  

We now discuss two simple models of the disk. Consider first a disk of uniform density $n$ (see Supplementary Information). 
The scale height of such a disk is $H/R \sim 0.01 \left( \frac{T}{10^4 \, \K} \right)^{1/2},$ 
implying that the disk is thin, and the values of the density and mass are
\be
	&& n =1.5 \times 10^5  \left( \frac{T}{10^4 \, \K} \right)^{-1/4} \left(\frac{R_\max}{ 9\, \mathrm{mpc}}\right)^{-3/2} \left( \frac{\M}{100}\right)^{-1/2} ~  \mathrm{cm^{-3}}\\
	&& M= 1.0 \times 10^{-4} \left( \frac{T}{10^4 \, \K} \right)^{1/4} \left(\frac{R_\max}{ 9\, \mathrm{mpc}}\right)^{3/2} \left( \frac{\M}{100} \right)^{-1/2} ~ M_{\odot}.
\ee
We estimate the rate of $10^4$ K gas accretion onto Sgr~A* at
\be
	\dot{M} (R_\mathrm{mean} )= 2.7 \times 10^{-10}  \left( \frac{T}{10^4 \, \K} \right)^{5/4} \left(\frac{\a}{0.1}\right) \left( \frac{\M}{100} \right)^{-1/2}  ~  {M_\odot}\,{\mathrm{yr}^{-1}},
\ee
where $\alpha$ is the dimensionless Shakura-Sunyaev viscosity parameter, and $R_\mathrm{mean} =(R_\max+R_\min)/2= 0.15 \, \arcsec.$
This is consistent with the upper limit on $\dot{M}$ from the lack of Faraday depolarization of Sgr~A* \cite{Bower03}, but lower than $\dot{M}$ predicted in $T \geq 10^7 \, \K$ gas at this radius in most radiatively inefficient accretion flow models \cite{Mahadevan98,Mahadevan97,Tchekhovskoy12,Yuan14}. 

A second model (see Supplementary Information) is similar to the broad-line region of quasars and Seyfert galaxies. In this model the observed emission may be distributed over many small $10^4$~K cloudlets, which might form in a thick torus. 
Disk accretion could be due to the cloudlets collisions, in which they transport angular momentum outward and rain onto the black hole at a rate of
\be
    \dot{M}(R_\mathrm{mean} )= 1.2 \times 10^{-7} \left( \frac{\x}{1/4} \right)  \left( \frac{n}{10^6 \, \cm^{-3}} \right)^{-3} \left( \frac{r}{9 \, \mathrm{\m pc}} \right)^{-1} \left(\frac{R_\max}{ 9\, \mathrm{mpc}}\right)^{-7/2}  \left( \frac{\M}{100}\right)^{-2} ~  {M_\odot}\,{\mathrm{yr}^{-1}},
\ee
where $r$ is the characteristic radius of a cloudlet, and  $n$ is the density within the cloudlet, and $\x=1/4$
reflects the fact that the peaks of the emission are near $\sim 500 \, \kms$ rather than at the
Keplerian velocity at the median radius of the disk, which is $\sim 2, 000 \, \kms.$ 
The mass of such a disk is
\be
	M=1.5 \times 10^{-5} \left( \frac{\M}{100}\right)^{-1} \left( \frac{n}{10^6 \, \cm^{-3}}\right)^{-1} \, \Msun.
\ee	
The disk therefore has to be replenished with $10^4 \, \K$ gas from the cooling of colliding winds and/or from the circumnuclear torus and the mini-spiral in order to exist on timescales beyond $\sim 120$ years. Note, the accretion rate and the lifetime estimates depends on the assumed density within cloudlets, disk size and the value of the amplification factor $\M,$ see Supplementary Information for the detail and derivations.
We modelled the geometry and dynamics of the gas emission assuming a broad line region disk \cite{Pancoast14a, Pancoast18} (see Supplementary Information). 

Assuming either of the above models, the disk is so tenuous, that the loss of the orbital angular momentum of the G2 envelope during the encounter should be minimal ($\lesssim 0.02$), and within the uncertainties of determining G2's orbital parameters \cite{Gillessen18}.

The result presented here is the consequence of pushing limits of ALMA. The line to continuum ratio is 1:2000, and an unprecedented accuracy of the continuum subtraction is required. We believe that we succeeded in achieving this using non-standard data analysis described in detail in the Supplementary Information section. Still, the possibility remains that the double-peaked line profile we see is an unlucky combination of molecular lines or is partially due to a foreground absorption. However, the spatial regularity of the integrated blueshifted and redshifted emission (Fig. 1bc), the fact that the width of the line emission is similar to the Keplerian velocity, the consistency of the ionization equilibrium calculations with the expectations,  and the very weak influence on G2-object orbit solidify out interpretation of the observed double-peaked profile as the H$30\alpha$ recombination line emission from a rotating a $10^4$ K disk within $\sim 2 \times 10^4 R_\mathrm{Sch}$ around Sgr~A*. 

New high resolution observations of Sgr~A* and its surroundings are transforming our view of the interaction of this prototypical slowly accreting, massive black hole with its environment. Observations with ALMA, GRAVITY and the Event Horizon Telescope and the upcoming Extremely Large Telescope and the Thirty Meter Telescope are complementary to each other and will connect the many currently puzzling views of Sgr~A*'s accretion flows, and clarify the radial dependence of inflow and outflow in different gas phases.

The 1971 paper by Lynden-Bell and Rees first predicted that the Galactic Center contained a massive black hole \cite{LR71}. The first of the ``critical observations'' which they suggested to test their hypothesis was of radio recombination lines, which ``should be $500 \, \kms$ broad and should shift by its own width between the two sides of the center.'' We are pleased to have confirmed this, after only 48 years.  

\section*{Acknowledgements}
We are grateful to Nick Scoville for co-writing the observing proposal and his contribution to discussions of analysis and interpretation of the data and to Jin Koda and Juergen Ott for discussing the data analysis, looking at the data and commenting on the paper.

We are grateful to Anna Ciurlo, Andrea Ghez, Mark Morris, and Stefan Gillessen for calculating the limit on $\Brg$ and sharing it with us and discussions, and to Eliot Quataert, Sean Ressler, and Jessica Lu for bringing the $\Brg$ non-detection to our attention and discussions. 
We would like to thank Yuri Levin, Jorge Cuadra, Peter Goldreich,  Doug Lin, James Guillochon, Smadar Naoz, Gunther Witzel, Sasha Philippov, Matt Coleman, Scott Tremaine, Avi Loeb, Zoltan Haiman, and John Carlstrom for discussions and comments, and to NAASC scientists at NRAO and in particular Tony Remijan, Catherine Vlahakis, Mark Lacy, Sabrina Stierwalt, Arielle Moullet, Erica Keller and Brian Kirk for their help and advice with the observational setups and data reduction.
We are grateful to Zara Scoville for proofreading the manuscript.

E.M.M. acknowledges the Bezos Fund for providing her stipend at IAS;
Dr. David and Barbara Groce for their encouragement and for supporting her as a Groce Fellow at Caltech; and 
NRAO Student Observational Support program for supporting 2 years of her graduate studies.

A.P. is supported by NASA through Einstein Postdoctoral Fellowship grant number PF5-160141 awarded by the Chandra X-ray Center, which is operated by the Smithsonian Astrophysical Observatory for NASA under contract NAS8-03060.
 
  ALMA is a partnership of ESO (representing
  its member states), NSF (USA) and NINS (Japan), together with NRC
  (Canada) and NSC and ASIAA (Taiwan) and KASI (Republic of Korea), in
  cooperation with the Republic of Chile. The Joint ALMA Observatory is
  operated by ESO, AUI/NRAO and NAOJ.
  
The National Radio Astronomy Observatory is a facility of the National Science Foundation operated under cooperative agreement by Associated Universities, Inc.

\newpage

\section*{Supplementary Information}

\renewcommand\thefigure{\arabic{figure}}    
\setcounter{figure}{0}

\subsection*{Observations}

The ALMA Cycle 3 observations of H30$\alpha$ line emission were obtained in April and August, 2016 for project 2015.1.00311.S.  Observations were centered on Sgr A*: RA 17:45:40.0359, DEC -29:00:28.169 (J2000).
They were conducted in receiver Band 6; the correlator was configured in the time division mode (TDM) with 4 spectrometers. Each spectrometer had a full bandwidth of 1875 MHz with 31.25 MHz resolution spectral channels. Because the width of the line was comparable to the bandwidth of the individual spectrometers, we overlapped two spectral windows centred at 231.058 GHz and 232.608 GHz (Supplementary Fig.~1). The remaining two spectral windows centred at 217.801 and 215.801 GHz were used to image the SgrA* continuum.

The observations were done primarily in ALMA configuration C40-5 with baselines up to 1.1 km and one execution was done in configuration C36-2/3 with baselines up to 460 m. For the C40-5 telescope configuration, good flux recovery is expected out to scales of $\sim$ 3.4 arcsec and for C36-2/3 it is expected up to $\sim$ 10.7 arcsec. Extended emission with spatial size greater than this will be partially resolved out. The data were taken with 43 12m antennas, using the total of 5.1 hours ALMA time, including calibrations. The integration time on target was 1.6 hours. The synthesized beam size was $0.29 \times 0.22 \, \arcsec^2$ with PA$=6$ degrees.
J1924-2914 was used as a bandpass calibrator. J1744-3116 or J1717-3342 were used as phase calibrators. J1924-2914 or J1733-1304 were used as flux calibrators.
The 1$\sigma$ (rms) sensitivity was 0.3 mJy beam$^{-1}$ in each 40 km s$^{-1}$ channel.
The observed line width was $\Delta V = 2,200 \, \mathrm{km \, s^{-1}}$,
 with a velocity resolution of $40 \, \mathrm{km \, s^{-1}}.$ 

Following delivery of data products, the data were re-reduced and imaged using the Common Astronomy Software Applications package (CASA). We did not perform self-calibration to preserve astrometry.  The analysis was performed in Python and Mathematica. 

The velocities given here are $V_\mathrm{radio} = c\frac{\nu_\mathrm{rest}-\nu}{\nu_\mathrm{rest}}$ relative to the LSRK. The Sgr~A* observations were centred on $V_\mathrm{LSR}=0.0 \, \mathrm{km \, s^{-1}}.$

\subsection*{Continuum Subtraction}

The continuum subtraction is the crucial part of the analysis. Here we go through it in detail.

We observed a $2,200 \, \mathrm{ km \, s^{-1}}$ wide line emission with mean flux $1.7$ mJy on top of the $\sim~2.8 - 3.5$ Jy, primarily from variable Sgr~A* synchrotron continuum. 
To achieve this we needed excellent bandpass calibrations, which were repeated at least every 30 minutes.

Spectral setup: The width of the observed recombination line is comparable with the width of the single spectral window. Therefore, the line had to be observed in two spectral windows to achieve secure coverage of the full line width and quality continuum subtraction. 

\begin{Supplementary Figure}
\centering
\includegraphics[width=8.9cm]{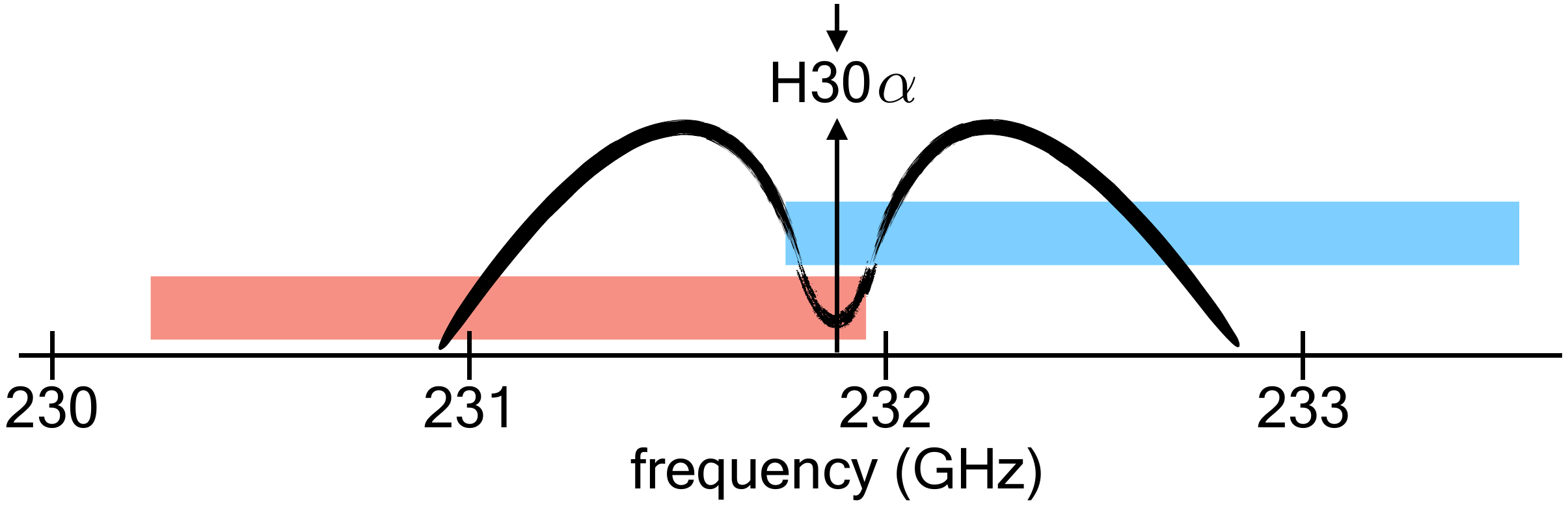}
\caption{
\textbf{Spectral windows configuration for our ALMA Cycle 3 observations.}
The observations have two spectral windows (blue and red) positioned across the line with an overlap. 
The frequency of the H30$\alpha$ line is marked with an arrow. The double peaked line between 231 GHz and 233 GHz is a schematic drawing of the detected line.
}
\label{spws}
\end{Supplementary Figure}

Spectral windows misalignment: The ALMA spectral windows are calibrated separately, and there is usually a small misalignment between the absolute value and the spectral slope of the  
data taken with one spectrometer (i.e. in one spectral window) with respect to the other. To ensure that separate spectral windows were consistently aligned, we 
 positioned spectrometers with overlap.  The overlap was established using $35$ side channels in each spectral window. Because $\sim10$ channels at the end of each spectral window are unusable this left us with $\sim 15$ useful channels to establish the proper alignment. 
The misalignment of the spectral windows in the delivered ALMA data and  the necessity to establish the proper spectral window alignment during the data analysis require working with each spectral window separately. 

Continuum subtraction procedure:
\begin{itemize}

\item Firstly, we imaged the data without continuum subtraction. The images were produced with the parameter Robust = 0.5. We cleaned images and identified the channels free of visible line features. 

\item Secondly, we performed u-v continuum subtraction using the identified channels, imaged the data again, and marked the channels with weak line features, which were not identified in the first step, to obtain a more accurate set of line free channels. 

\item Thirdly, we repeated the second step as many time as was necessary to identify all visible line features and achieve good continuum subtraction. 

\item When steps one through three are performed correctly in each spectral window the resulting continuum subtracted spectra will be aligned, i.e. the overlapping channels in each spectral window will match. In this work we achieved the alignment  of the spectral windows with the u-v continuum subtraction procedure outlined above. 

\end{itemize}

It is extremely important to establish the alignment of the separate spectral windows. Misaligned spectral windows will render further analysis nonsensical. However, the procedure described above is very time-consuming. The observed recombination line is wide which leaves a small number of channels grouped in a narrow spectral range at one of the ends of the spectral window for continuum subtraction. The channels have noise and the average obtained from them is not always accurate when interpolated across the whole spectral range. Adding or subtracting one channel to the list of ``line-free channels'' may affect the slope or the quality of the continuum subtracted spectrum. 

A quicker procedure may be used to align the spectra instead. One may identify a reasonable list of the line-free channels (we quantitatively define this below), do u-v continuum subtraction, clean the image with the CASA task TCLEAN, extract spectra in a selected aperture in each spectral windows separately, and align the parts of the spectra by subtracting a straight line $a \nu + b,$ with $a$ and $b$ constants, to match the overlapping parts of the spectra. (Due to the spectral window misalignment, the correct subtraction required different values $a$ and $b$ in each spectral window.) 

We experimented with this procedure. We defined a reasonable list of the line-free channels to be one resulting in slopes of no more than $< 30$ mJy over the width of the spectral window after the u-v continuum subtraction.  (Large slopes result in larger scatter of the points in the spectra. The trend observed in this experiment is that the task TCLEAN builds up the points at the top of the slope making them even higher.)  After aligning the overlapping parts of the spectra we compared the join spectra obtained after u-v continuum subtraction with different sets of ``line-free channels''. The variations in the value of each point in the resulting line spectra were $\leq 7\%.$  Therefore to those willing to undertake re-analysis of our data we recommend this procedure instead of the more tedious approach we took ourselves.

Regarding the identification of line-free channels, it is important to stress that the Galactic Center is a complex region containing stars, clouds and streamers moving with various velocities and therefore molecular line features are shifting in the velocity space from one spatial location to the other. This render identification of a unique set of line-free channels applicable everywhere across the region impossible. This, in general, is not an issue for strong line features. However, in the case of weak emission, such as the one we discuss in this work, it poses a problem. The shifting molecular features have flux comparable to or stronger than the line itself, and when (partially) shifted into the identified set of the line-free channels would drastically influence the quality of the continuum subtraction. In this work, we identified the reasonable list of the line-free channels only within  $<1$ arcsec around Sgr A*.
We tested it within the aperture comparable to the beam size.

\subsection*{Reliability}

To further test the reliability of our results we performed the following:
\begin{itemize}

\item We split the data into separate observation blocks (separate ALMA executions made on different days) and processed them independently without reference to each other. In each execution the value of the continuum was different and varied from $\sim 2.8$ Jy to $\sim 3.5$ Jy. 
In all cases we recovered the similar line shape, spectral width, the peak recombination line flux, and integrated line flux, excluding the possibility of the calibration artifact, as the calibration artifact would have a multiplicative effect on the observed spectra;

\item We observed a continuum source J2000-1748, 
which was calibrated in the same way as Sgr~A*, to check for possible technical or data reduction errors.
No spectral feature analogues to those in Fig.~1 were detected, i.e. no double peaked line with wings of the width $\sim 1000 \, \kms$;

\item We analyzed ALMA Cycle 4 observations of Sgr A* conducted a year later on the same line, but with different spectral windows setup. All four spectra windows were positioned in one side band and with the overlap of $\sim 1/3$ of the spectra window width. There are three spectral windows across the observed recombination line). We recovered the similar line shape, spectral width, the peak recombination line flux, and integrated line flux. The integrated line flux was larger by $10 \%$ compare to the Cycle 3 data presented here. Joint analysis of ALMA Cycle 4 and Cycle 5 observations will be published separately;

\item Our ALMA Cycle 4 observations also had a continuum source --  J1752-2956 (different from the one in the Cycle 3 observations), which was calibrated in the same way as Sgr~A*, to check for possible technical or data reduction errors.
No spectral feature analogues to those observed on Sgr A* spectra were detected.

\end{itemize}

\subsection*{Estimation of Uncertainties}

\begin{itemize}

\item The extent of the emission is determined by analyzing the spectra within different apertures. An increase in the aperture size beyond $0.63 \times 0.51 \, \arcsec^2$ does not increase the H30$\a$ flux within it. Emission outside of this region, if any, is $<10\%$ of that inside.

\item  The observational sensitivity is $\delta S = 0.3$ mJy in each 40 $\mathrm{km \, s^{-1}}$ channel. Given that the mean value of the detected line is 2.5mJy, this gives us 13\% uncertainty.

\item We estimate the average uncertainty due to variations in u-v continuum subtraction and  subsequent TCLEAN application at $5 \%$ on average (see section Continuum Subtraction).

\item The overlapping parts of the spectra used to ensure proper spectral windows alignment do not match perfectly. They allow for variation in the alignment, specifically in the slopes of the aligning parts of the spectra (see section Continuum Subtraction).
We estimate that the alignment uncertainty may result in $\sim 10\%$ variation of the velocity integrated line flux.

\item  The molecules with lines within $\sim 1$ GHz around H30$\a$ are acetone, methanol, sulfur dioxide $^{33}SO_2,$ and similar complex molecules \cite{splatalogue16}.
They are not expected to be 
present in substantial quantities in $10^4 \, \K$ gas.
However, it is hard to exclude the possibility of narrow absorption or emission features from the foreground. 
The spectrum shows a relatively narrow $150 \, \mathrm{km \, s^{-1}}$ bump at 231.43 GHz, which might be due to foreground emission. This feature is responsible for $0.2 \, \mathrm{Jy \, km \, s^{-1}}$ in the total velocity integrated line flux of $S\D V_{\mathrm{H}30\a}=3.8 \, \mathrm{Jy} \, \kms.$

\item We explore a relatively narrow spectral range of frequencies, while the wings of the line might extend further than $\pm 1000 \, \mathrm{km \, s^{-1}}$ from the 
central frequency.  
 In these observations we are unable to explore smooth and extended line features comparable in the velocity width to the width of the spectral window $\sim 2000 \, \kms.$
This will be tested in our ALMA Cycle 5 observations which will be conducted in the spectral scan mode and cover $20,000 \, \mathrm{km \, s^{-1}}$. 

\item We have expected higher noise than the one we are seeing in the spectra Fig.~1. 

\item The sharp dip at 230.9 GHz is robustly detected. So is the blue shifted wing of the line, which clearly rises from the flat continuum identified as the range between $-2000 \, \kms$ and $-1200 \, \kms.$
The position of the redshifted wing is less certain due to the narrow frequency range available between the CO line and the end of the band. The redshifted wing is positioned in place with the help of the overlapping spectral points which are mostly located in the dip.  

\item In the image made from the blue-shifted side of the spectrum,  a negative
intensity feature is present at the location of the reshifted emission. We will investigate this in our further data analysis with new data. 

\item It is hard to exclude the possibility that the dip in the middle is the result of the foreground absorption. There is a possible extended absorption feature at $\sim 230.9$ GHz to the West from Sgr A*. We will investigate this in our further data analysis with new data. A possibility also remains that the double-peaked line profile we see is noise or a combination of molecular lines. However, the combination of the following facts: that a similar spectral structure is detected both in ALMA Cycle 3 and in Cycle 4 (to be published separately), that regular structure is seen in the integrated blueshifted and redshifted emission, and that the width of the line emission is consistent with Keplerian velocity at the detected radius, makes us conclude that the chances of such a spectral structure appearing coincidentally at this frequency is quite low. 

\end{itemize}

Combining the above factors we estimate that the combination of the listed above factors gives a combined uncertainty of $\sim 20 \%$ of the value of each point on the presented continuum-subtracted line emission spectra, and an additional 10\% uncertainty to the velocity integrated line flux due to the alignment of the separate spectral windows during which the curved representing the wings of the spectrum shift as the whole.  

\subsection*{Accretion onto Sgr A*, an overview.}

A supermassive black hole, Sagittarius A*, lies at the center of our Galaxy. It has mass $M_\mathrm{SgrA*} \simeq 4 \times 10^6 M_{\odot}$  {\cite{Boehle16,Gravity18}}.  
Building up such a massive black hole within the $\sim 10^{10}$ year lifetime of the Galaxy would require a mean accretion rate of $\sim 4 \times 10^{-4} \, \Msun \, {\yr}^{-1}.$ 
However, polarization measurements constrain the rate of gas accretion near the event horizon $R_s$ to $\dot{M}_\mathrm{horizon} \sim 10^{-9}-10^{-7} M_{\odot} \, \mathrm{yr}^{-1}$ \cite{Bower03}, and X-ray observations constrain it at the Bondi radius ($10^5 \, R_\Sch = 0.04 \, \pc$)  to
 $\dot{M}_\mathrm{Bondi} \sim 3 \times 10^{-6} M_{\odot}\,\mathrm{yr}^{-1}$ \cite{Baganoff03,Quataert02,Quataert04}.
If the radiative efficiency were $\sim 10\%,$ the $\dot{M}_\mathrm{Bondi}$ would yield a luminosity $\sim 10^4$ times Sgr~A*'s bolometric luminosity of 
a few $ 10^{36} \, {\textrm{erg} \, \textrm{s}^{-1}}$  {\cite{Narayan98}}.

Extensive theoretical efforts have been put into resolving the mystery of Sgr~A* accretion, and Radiatively Inefficient Accretion Flows (RIAF) in general.
The models describe an accretion disk which cannot efficiently cool \cite{Phinney81,BBRP82} and is geometrically thick.
The macroscopic effects transfer energy primarily to the ions. The ions lose only a small fraction of their energy to the electrons through Coulomb scattering on an inflow/heating timescale. As a result, the radiation efficiency of such a flow is very low, and the gas falling into the horizon radiates only $\ll 0.1 \dot{M} c^2$ (a radiatively efficient thin accretion disk radiates  6\% - 42\% of $\dot{M}c^2$ depending on the BH spin) \cite{Thorne74,Novikov.Thorne73},  allowing the observed luminosity to be produced by a much higher accretion rate than in a thin disk, 
while also explaining the mm to $\gamma-$ray spectrum \cite{Narayan.Yi.Mahadevan95}.

An Advection-Dominated Accretion Flow (ADAF) \cite{Narayan.Yi95} resembles a thick disk and rotates at an angular velocity much less than the Keplerian velocity $\Omega \ll \Omega_\mathrm{K}.$ The small amount of radiation loss (the amount of energy transferred from ions to electrons) is estimated from Coulomb 
collisions, or set to be a free parameter to account for plasma effects. The black hole is fed at a constant rate and no material escapes. The density of such a disk scales as $\rho \sim r^{-\frac{3}{2}}.$ 
However, an ADAF may be unstable to driving a wind. 
An Advection Dominated Inflow Outflow Solution (ADIOS) \cite{Blandford.Begelman99}  is characterized by the presence of both an inflow and an outflow. It has the geometrical characteristics of the ADAF solution, but
the disk has an accretion rate decreasing with radius as a power law, as the winds blow away material in the outer parts of the disk. The density profile $\rho \sim r^{-\frac{3}{2}+p},$ where $p$ is a constant parameter, is  less steep than the ADAF, and the absolute value of the density is lower.
ADAFs may be unstable to convection, so a Convection-Dominated Accretion Flow (CDAF) was proposed \cite{Quataert.Gruzinov00}. This accretion flow is marginally stable when the convection dominates advection in carrying the material inwards. A CDAF is also a thick disk rotating at a much lower angular velocity than Keplerian velocity, and feeds the black hole at a constant rate, but the density of such a flow scales as $\rho \sim r^{-\frac{1}{2}}.$ 
We should also mention that in numerical simulations which included magnetic fields and a jet, it was obtained that $\rho \sim r^{-1}$ \cite{Tchekhovskoy12}.

A detailed fit of no-wind ADAF models to the observed Sgr~A* spectra from radio to $\gamma$-rays  \cite{Mahadevan97,Mahadevan98} led to the estimate of the black hole accretion rate at
\be
	\dot{M}_\mathrm{SgrA*} = 7 \times 10^{-6} \left( \frac{\a}{0.3} \right) \frac{M_{\odot}}{\mathrm{yr}},
\ee
where $\alpha$ is the dimensionless Shakura-Sunyaev viscosity parameter \cite{Shakura.Sunyaev73}.
The no-wind ADAF causes a pile up of material in the accretion zone such that it becomes inconsistent with Faraday rotation measurements \cite{Bower03,Aitken00,Agol00,Marrone07}. 
Assuming that the magnetic field is ordered and at equipartition strength, the rotation measure constrains the accretion rate to a much lower value of
\be
	\dot{M}< 2 \times 10^{-7} \, \frac{M_{\odot}}{\mathrm{yr}},
\ee
though the assumptions make this constraint rather model dependent.

Inclusion of an outflow solves the pile-up issue. 
The detailed fit of RIAF models with an outflow to  the spectrum of Sgr~A* from radio to $\gamma$-rays \cite{Yuan14}
results in
\be
	&& \dot{M}_\mathrm{Bondi} \sim 3 \times 10^{-6} {M_{\odot}}/{\mathrm{yr}} \\
	&& \dot{M}_\mathrm{SgrA*} = 1.2 \times 10^{-7} {M_{\odot}}/{\mathrm{yr}},
\ee
which is consistent with the constraint from the Faraday rotation measurements.
There is no observational evidence for the presence of an outflow near Sgr~A*. There is, however, no evidence excluding such a possibility either.
A recent hydrodynamic simulations of the inner accretion flow of Sgr~A* fueled by stellar winds obtained \cite{Ressler18}
\be
	\dot{M}_\mathrm{SgrA*}=2.4 \times 10^{-8} \frac{M_{\odot}}{\mathrm{yr}} \left( \frac{R}{R_\Sch} \right)^{1/2}.
\ee

It has been difficult to favor or rule out any of these accretion models for Sgr~A*, primarily due to the lack of model-independent
observational constraints on the accreting gas behavior between $10$ and $10^5 \, R_\Sch.$ 
We would like to stress that we are talking about the constraints which do not rely on assuming particular scaling of the accretion flow properties with radius to make a prediction. Constraints on the accretion rate from the Faraday rotation measurements \cite{Bower03,Aitken00,Agol00,Marrone07} are model-dependent as they assume scaling of density and the magnetic field strength with radius as input parameters. Constraints on the average density of the accretion flow from the drag on the G2-object's orbit \cite{Gillessen18} employs the density scaling with radius of the accretion flow as $\sim r^{-1}$ and assumes the flow is not rotating, and thus it is also model-dependent.

In this work we constrain the quantity and the dynamic properties of the cool $T\ll 10^7 \, \K$ gas in the accretion zone of Sgr~A*. 
We would like to emphasize that although our estimations discussed in details below depend on the model of the cool disk, they do not assume a model for the accretion flow or the scaling of its parameters with radius.

\subsection*{Velocity Integrated Line Flux and Volume Emission Measure.}

\begin{Supplementary Figure}[p]
\centering
\includegraphics[width=15cm]{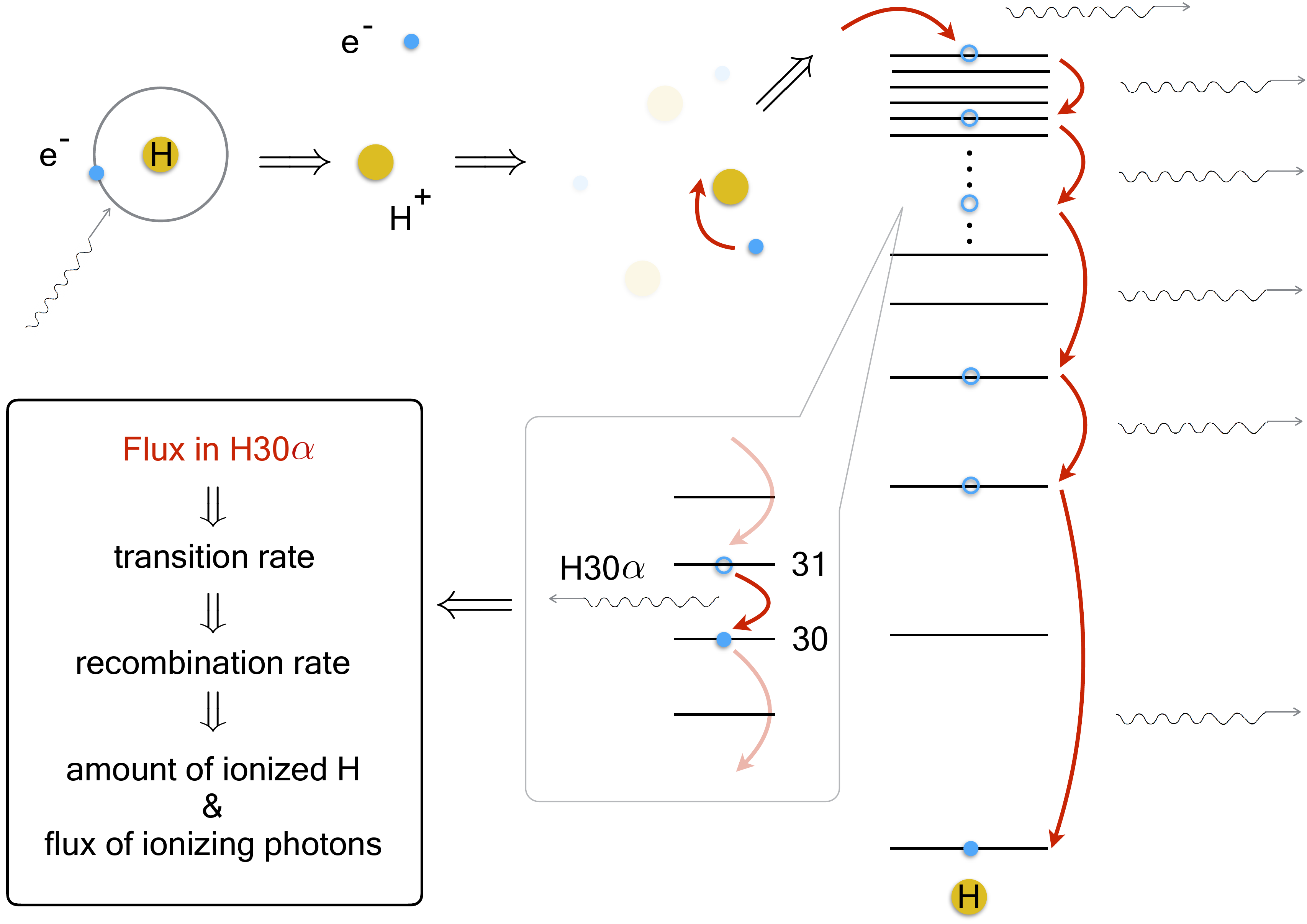}
\caption{
\textbf{Schematic plot illustrating the recombination line technique.} If photons with energy $E_\gamma \geq 13.6$ eV are present, they ionize neutral hydrogen. As the electrons and protons recombine, some recombinations occur to $\mathrm{n}\gg 1.$ The electrons cascade down to the ground level. Some electrons pass through the levels of interest, in this case H$30\alpha:$  $\mathrm{n} =31 \to 30,$ during the cascade. The amount of radiation coming out in H$30\alpha$ indicate how many H30$\alpha$ transitions are occurring, how many atoms are recombining, how much ionized material is in the region, and the background flux of the ionizing photons.
}
\label{recombtech}
\end{Supplementary Figure}

In this section, we treat recombination line emission in the conventional case of no background pumping, so no maser emission.  The main text discusses the evidence from Br~$\gamma$ limits for masing in the H30$\alpha$ line, and the resulting correction factor $\M$ to $\epsilon_{\rm H30\alpha}$ and the density inferred from the emission measure.

Supplementary Fig.~2 shows a schematic of the recombination line technique.
The H30$\a$ line luminosity is given by an integration over the line emitting region
\be
    L_\mathrm{H30\alpha}=\int \epsilon_{\textrm{H}30\a} n_e n_p d^3 r,
\ee
where $\epsilon_{\mathrm{H}30\a} $ is the emissivity of H$30\alpha$, which is a function of density and temperature, $n_e$ is the electron number density,
$n_p$ is the proton number density, and $d^3 r$ is the three dimensional integral over the emitting volume.
The flux received by the telescope is 
\be
    S_\mathrm{H30\alpha} = \frac{L_\mathrm{H30\alpha}}{4 \pi D^2}= \frac{\int \epsilon_{\textrm{H}30\a} n_e n_p d^3 r}{4 \pi D^2}.
\ee
Here $D$ is the distance to the emitting source.

In realistic cases, this flux is spread over a range of frequencies due to motion within the gas. The integrated line flux is then
\be
    \int\limits^{\nu_\max}_{\n_\min} S_\nu d\nu =\left( \int^{V_\max}_{V_\min} S_\nu dV \right) \frac{\nu_{\textrm{obs}}}{c} = (S\D V_{\textrm{H}30\a}) \frac{\nu_{\textrm{obs}}}{c}=  \frac{\int \epsilon_{\textrm{H}30\a} n_e n_p d^3 r.}{4 \pi D^2},
\ee
where $S_\nu$ is the line flux per unit frequency, $S\D V_{\textrm{H}30\a}$ is the velocity integrated line flux, $c$ is the speed of light, and $V$ is the line of sight velocity corresponding to the observed frequency's Doppler shift from the rest frequency of the H30$\a$ line. We find
\be
    S\D V_{\textrm{H}30\a}  = \frac{\int \epsilon_{\textrm{H}30\a} n_e n_p d^3 r}{4 \pi D^2} \frac{c}{\nu_{\textrm{obs}}}.
\ee
In a simplified case of an emitting region of constant density and temperature, this equation is reduced to
\be\la{SdVeq}
	S\D V_{\textrm{H}30\a}  = \frac{\epsilon_{\textrm{H}30\a} n_e n_p \vol }{4 \pi D^2} \frac{c}{\nu_{\textrm{obs}}},
\ee 
where $\vol = \int d^3 r$ is the total volume in space occupied by the $10^4 \, \K$ gas.
The emissivity $\epsilon_{\textrm{H}30\a}(T,n)$ varies weakly with density $n$ \cite{Storey.Hummer95}: 
\be
&& {\epsilon_{\textrm{H}30\a}(10^4 \, \textrm{K}, 10^4 \, {\mathrm{cm^{-3}}})} = 1.05 \times 10^{-31} \, \textrm{erg} \, \textrm{s}^{-1} \textrm{cm}^3, \\
&& {\epsilon_{\textrm{H}30\a}(10^4 \, \textrm{K}, 10^5 \, {\mathrm{cm^{-3}}})} = 1.08 \times 10^{-31} \, \textrm{erg} \, \textrm{s}^{-1} \textrm{cm}^3, \\
&& {\epsilon_{\textrm{H}30\a}(10^4 \, \textrm{K}, 10^6 \, {\mathrm{cm^{-3}}})} = 1.25 \times 10^{-31} \, \textrm{erg} \, \textrm{s}^{-1} \textrm{cm}^3, \\
&& {\epsilon_{\textrm{H}30\a}(10^4 \, \textrm{K}, 10^7 \, {\mathrm{cm^{-3}}})} = 1.36 \times 10^{-31} \, \textrm{erg} \, \textrm{s}^{-1} \textrm{cm}^3. 
\ee
In what follows we assume
\be\la{eps106}
	\epsilon_{\textrm{H}30\a} \simeq \epsilon_{\textrm{H}30\a} (10^4 \, \textrm{K}, 10^6 \, {\mathrm{cm^{-3}}}) .
\ee

Substituting equation \ref{eps106}, $\nu_\textrm{obs}=\nu_{\textrm{H}30\a} = 231.9$ GHz, $D_\mathrm{H30\alpha} = 8.0$ kpc, and $1 \, \textrm{Jy} = 10^{-23} \frac{ \textrm{erg} }{ \textrm{s}\, \textrm{cm}^{2} \, \textrm{Hz}}$ into equation \ref{SdVeq} we find
\be
    S\D V_{\textrm{H}30\a} = 2.1 \times 10^{-60}  n_e n_p \vol \,\, \textrm{Jy} \, \textrm{km} \, \textrm{s}^{-1}
\label{sdvf}
\ee
and the expression for the volume emission measure is
\be
    \textrm{EM} = n_e n_p \vol=S\D V_{\textrm{H}30\a} \times 4.7 \times 10^{59} \,\, \cm^{-3}.
\ee

\subsection*{Disk of a uniform density}

Let us consider a Shakura-Sunyaev disk model. For simplicity we assume $n=n_e = n_p= \const.$
The disk properties are as follows: 
an isothermal disk at $T=10^4~\mathrm{K}$ with an outer radius $R_\mathrm{max} = 0.23$ arcsec and an inner radius $R_\mathrm{min} = 0.07$ arcsec,  
a half opening angle $\phi,$ such that $H/R = \tan{\phi},$ where $H$ is the scale height of the disk measured from the midplane to the top. 
The disk rotates with an azimuthal velocity $V_\Omega$ equal to the Keplerian velocity $V_\mathrm{K}$  $(V_\Omega = V_\mathrm{K}).$

The emission measure is given by
\be\label{emc}
	\textrm{EM} = n^2 \frac{4}{3}\pi R_\max^3  \tan{\phi} \left[ 1-\left( \frac{R_\max}{R_\min} \right)^3 \right].
\ee
Making use of equation \ref{em} of the main text and its preceeding paragraph's
definition of the factor $\M$ to account for possible masing, we find
\be
	&& n = \frac{1.5 \times 10^4}{\sqrt{\tan \phi}} \left(\frac{R_\max}{ 9\, \mathrm{mpc}}\right)^{-3/2} \left( \frac{\M}{100}\right)^{-1/2} ~  \mathrm{cm^{-3}},\\
	&& M= 1.0 \times 10^{-3} \sqrt{\tan \phi} \left(\frac{R_\max}{ 9\, \mathrm{mpc}}\right)^{3/2} \left( \frac{\M}{100}\right)^{-1/2} ~ M_{\odot}.
\ee

Then the mass accretion rate onto the black hole $\dot{M}$ is 
\be
	\dot{M} = 2 \pi R 2 H(R) V_{R} \rho(R), 
\ee
where $V_{R}$ is the radial inflow velocity of the gas at the radius R and $\rho=n m_p$ is the gas mass density.
The expression for the radial inflow velocity in the accretion disk is
\be\label{rel}
	V_R= \alpha \left( \frac{H}{R} \right)^2 V_\Omega = \alpha \left( \frac{H}{R} \right)^2 \x V_K,
\ee 
where $c_s$ is the speed of sound and $\alpha$ is the dimensionless Shakura-Sunyaev viscosity parameter \cite{Shakura.Sunyaev73} and
$\x=V_{\Omega}/V_{K}$ is the parameter describing deviation of disk's material orbital velocity from Keplerian velocity.
We find that the accretion rate at the radius $R_\mathrm{mean}=(R_\max+R_\min)/2$ is
\begin{equation}
\begin{array}{l}
	\displaystyle \dot{M}=  4 \pi R^2 n m_p \x V_K \a \left( \frac{H}{R} \right)^3 \\
	\displaystyle \quad \,\, = 3.2 \times 10^{-5} \left( \tan{\phi} \right)^{5/2} \x \left(\frac{\a}{0.1}\right)  \left( \frac{\M}{100}\right)^{-1/2}    \frac{M_\odot}{\mathrm{yr}}.
\end{array}
\end{equation}

The scale height of the disk can be estimated using
\be
	\tan \phi = \frac{H}{R}= \frac{c_\textrm{s}}{V_{\Omega} }.
\ee
The speed of sound in the ideal gas is 
\be
	c_s =\sqrt{\frac{\gamma p}{\rho}} =\sqrt{\frac{\gamma (n_e +n_p)k_B T}{n_p m_p}}=\sqrt{\frac{\gamma 2 k_B T}{m_p}}= 16.6 \left( \frac{T}{10^4 \, \K} \right)^{1/2} ~ \mathrm{ km \, s^{-1}},
\ee
then
\be
	\tan \phi=\frac{c_\textrm{s}}{V_{\Omega} } \simeq \frac{16.6 ~ \mathrm{ km \, s^{-1}}}{2,000 ~ \mathrm{ km \, s^{-1}}}  \sim 0.01 \left( \frac{T}{10^4 \, \K} \right)^{1/2}.
\ee
Finally we have
\be
	&& n =1.5 \times 10^5  \left( \frac{T}{10^4 \, \K} \right)^{-1/4} \left(\frac{R_\max}{ 9\, \mathrm{mpc}}\right)^{-3/2} \left( \frac{\M}{100}\right)^{-1/2} ~  \mathrm{cm^{-3}}\\
	&& M= 1.0 \times 10^{-4} \left( \frac{T}{10^4 \, \K} \right)^{1/4} \left(\frac{R_\max}{ 9\, \mathrm{mpc}}\right)^{3/2} \left( \frac{\M}{100}\right)^{-1/2} ~ M_{\odot}\\
	&& \dot{M} = 2.7 \times 10^{-10}  \left(\frac{\x}{1}\right) \left( \frac{T}{10^4 \, \K} \right)^{5/4} \left(\frac{\a}{0.1}\right) \left( \frac{\M}{100}\right)^{-1/2}  ~  \frac{M_\odot}{\mathrm{yr}}.
\ee

A disk with these properties is gravitationally stable, since the Toomre $Q$ in the centre of the structure is much greater than one: $Q= \frac{V_\mathrm{K}/R c_s}{\pi G m_p n H} \sim 1.7 \times 10^7 \gg 1$ .

\subsection*{BLR-like Ensemble of Clouds.}

It is easier for a maser explanation
of the difference in emissivity inferred from these observations and
the preliminary Br$\gamma$ estimates if the beam filling factor
is low. This suggests consideration of a disk of cloudlets.

Let us consider a thick disk consisting of Broad-Line-Region (BLR)-like cloudlets.
The model is as follows: 
a disk with an outer radius $R_\mathrm{max} = 0.23$ arcsec and an inner radius $R_\mathrm{min} = 0.07$ arcsec,  
a half opening angle $\phi,$ such that $H/R = \tan{\phi},$ where $H$ is the scale height of the disk measured from the midplane to the top, filled with isothermal cloudlets at $T\sim 10^4$ K. The cloudlets have a characteristic radius $r,$ an internal density $n=n_e = n_p$ and they move in circular orbits  with velocities proportional to Keplerian velocities $V_\Omega=\x V_\mathrm{K}$ with $\chi \in (0, 1].$  
The volume of such a disk is 
\be
	\vol_\mathrm{disk} =\frac{4}{3}\pi \tan{\phi} R^3_\max \left( 1-\frac{R^3_\min}{R^3_\max} \right),
\ee
and the volume emission measure is
\be
	\EM = n^2 \times \vol_\mathrm{cloud} \times n_\mathrm{cloud} \vol_\mathrm{disk}.
\ee

The accretion is due to cloudlet collisions, during which they lose angular momentum and an amount of material, approximately equal to the mass of a cloudlet, ``rains'' down on the black hole.
The collision rate per cloud is 
\be
	z_\mathrm{cloud}= \pi r^2 \times \x \frac{H}{R} V_\mathrm{K} \times n_\mathrm{cloud} .
\ee
Here $n_\mathrm{cloud}$ is the number density of the cloudlets within the disk, and $\x \frac{H}{R} V_\mathrm{K}$ is the velocity of the clouds relative to each other.

The mass accretion rate for such  a disk is independent of its opening angle $\tan \phi:$
\be
	&& \dot{M}= m_\mathrm{cloud} \times z_\mathrm{cloud} \times  n_\mathrm{cloud} \vol_\mathrm{disk}  \\
	&&  \quad \,\, = \frac{\EM^2}{ \vol_\mathrm{disk}/\tan{\phi}} \frac{m_p \x V_\K}{4 r n^3}  \\
	&& \quad \,\, = {1.2 \times 10^{-7}} \left(\frac{\x}{1/4}\right)  \left( \frac{n}{10^6 \, \cm^{-3}} \right)^{-3} \left( \frac{r}{ 9 \mu \pc} \right)^{-1} 
	\left(\frac{R_\max}{ 9\, \mathrm{mpc}}\right)^{-7/2} \left( \frac{\M}{100}\right)^{-2} ~  \frac{M_\odot}{\mathrm{yr}}.
\ee
Here we used $r=10^{-3} R_\mathrm{max} = 9 \mu \pc$ as a characteristic size of cloudlets, $V_\K$ at the mean radius of the disk $R_\mathrm{mean} = (R_\max+R_\min)/2 = 0.15 \, \arcsec,$ 
and $\x = 1/4$ as it is the ratio between the velocity where the most emission comes from $V_\Omega \simeq 500 \, \kms$ and Keplerian velocity $V_\K\simeq 2,000 \, \kms$ at $R_\mathrm{mean}.$
The mass of this disk is
\be
	M_\mathrm{disk} = m_p \frac{EM}{n} = 1.5 \times 10^{-5} \left( \frac{\M}{100}\right)^{-1} \left( \frac{n}{10^6 \, \cm^{-3}} \right)^{-1}  ~ M_{\odot}.
\ee
The mass of the cloudlet is
\be
	m_\mathrm{cloud} = m_p n \frac{4}{3}\pi r^3 = 7 \times 10^{-11} \left( \frac{n}{10^6 \, \cm^{-3}} \right) \left( \frac{r}{ 9 \mu \pc} \right)^{3}  \, \Msun .
\ee
The number density of cloudlets in the disk is
\be
	n_\mathrm{cloud}  =  \frac{M_\mathrm{disk}/m_\mathrm{cloud}}{\vol_\mathrm{disk}} \simeq \frac{70}{\tan{\phi}} \, \mathrm{mpc}^{-3}.
\ee
The lifetime of the disk strongly depends on the assumed density within the cloudlets. For $n=10^6 \, \cm^{-3},$ the disk has to be replenished with $10^4 \, \K$ gas supplied by 
the circumnuclear torus and the mini-spiral and/or the cooling of colliding winds in order to exist beyond $\sim 120$ years. The gas supplied form by the cool gas structures around the Galactic Center \cite{Goicoechea18} has to survive the hot environment near the black hole without being complete evaporated. While hot gas has to be able to cool on the reasonable timescale.  The average density of the gas in the hot estimated accretion flow
is estimated at $\sim 10^2 - 10^3 \, \cm^{-3}$ at the radius of the disk \cite{Ressler18}. However if colliding stellar winds create overdensities $\sim 10^6 \, \cm^{-3},$ such clumps would cool from $10^7 \, \K$ on faster than the dynamic timescale up to the $R<R_\mathrm{disk}$.
Should the density within the clump be a little higher the lifetime could easily be 1000 year. For a long-lived disk there is of course a possibility that the disk was formed from a one-time infall event. 
However we think that replenishing is more likely a continuous process as the Galactic Center is a complex region with no shortage of gas supply either hot or cold. 

\subsection*{BLR-like Ensemble of Clouds. Modelling.}

If the gas disk is similar to the BLR in active galactic nuclei (AGN), we can model the geometry and dynamics of the gas emission using a method applied to the optical H$\beta$ broad emission line \cite{Pancoast14a, Pancoast18}.  
The model assumes an ionizing photon source located at the location of SgrA*, an outer radius of emission less than 10 light days (corresponding to the spectrum aperture), and a black hole mass of $4.0 \times 10^6 {M_\odot}$.  Fitting the H30$\alpha$ emission line to within the spectral uncertainties tightly constrains the BLR model parameters as shown in Supplementary Fig.~3.  The dynamics are dominated by outflowing orbits with more tangential than radial velocities, a result that is consistently inferred even if the black hole mass is left as a free model parameter or the outer radius of emission is larger than 10 light days.  The geometry parameters are much more sensitive to the value of black hole mass and the maximum radius of emission, due to degeneracies between the disk thickness, radial size, and orientation with the black hole mass.  The inferred geometry is a slightly thick disk viewed close to face-on and with a ratio of the mean radius to the minimum radius of emission inferred to be $1.78$.  Although the input to the modelling code was only the H30$\alpha$ line spectrum, it infers disk properties similar to the ones we deduce from the imaging.  These results also hint at an observational signature of outflow in the Galactic Centre. 

The red/blue shifted emission we observe might be due a bipolar outflow, rather than a rotating disk. This scenario is disfavored in our modelling, however it cannot be completely ruled out.
Future higher resolution observations could help to distinguish these cases.

\begin{Supplementary Figure}[p]
\centering
\includegraphics[width=15cm]{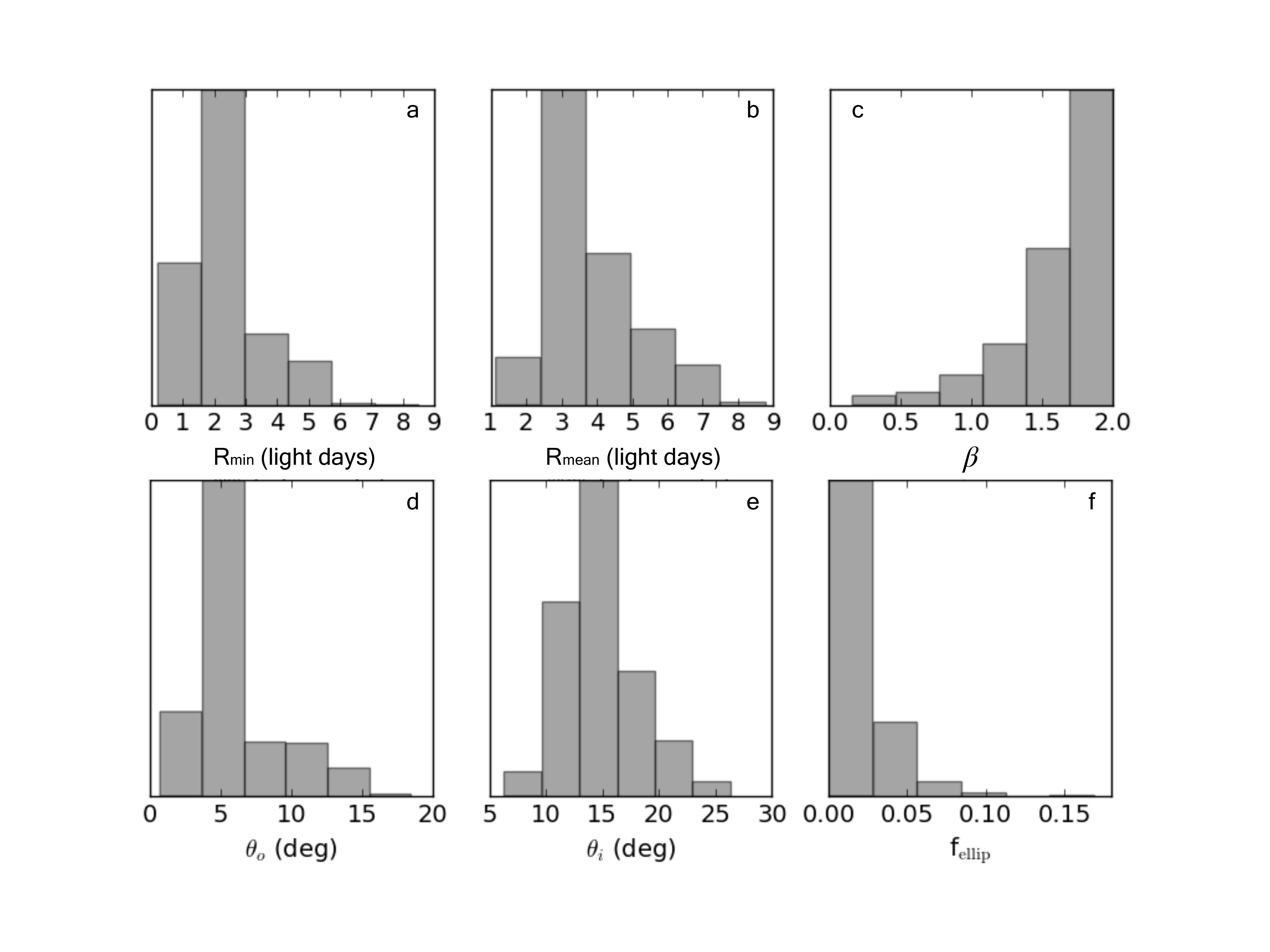}
\caption{
\textbf{Inferred posterior probability distributions for key BLR model parameters.}  The radial distribution of line emission is described by the three parameters in the top row, including (a) the minimum radius of emission $R_{\min}$, (b) the mean radius of emission $R_\mathrm{mean}$, and (c) the Gamma distribution shape parameter $\beta$, where the radial profile of emission is $\propto x^{1/\beta^2-1} \exp(-x)$.  Values of $\beta \to 0$ correspond to a narrow Gaussian-like radial profile, while $\beta = 1$ corresponds to an exponential profile and $\beta > 1$ (as preferred by the data) corresponds to a profile that decreases more steeply than exponentially with radius.  The median value and 68\% confidence intervals for the ratio of the mean radius to the minimum radius are $1.78^{+0.22}_{-0.30}$.  The opening angle $\theta_o$ (d) is slightly thick, where $\theta_o = 0$ (90) deg is a perfectly thin disk (sphere).  The inclination angle $\theta_i$ (e), at which an observer views the disk, is inferred to be close to face-on, where  $\theta_i = 0$ (90) deg is perfectly face-on (edge-on).  The fraction of line emission from gas on near-circular orbits (f), $f_{\rm ellip},$ is generally less than 10\%, with the remaining gas in outflowing orbits having predominantly tangential, instead of radial, velocities.  In addition to the posterior PDFs shown here, the geometry of emission is inferred to have some asymmetries in the angular coordinate direction, including more emission from the far side of the disk further from the observer and also more emission from above the disk mid-plane.  Combining these two forms of asymmetry, the brightest part of the disk is the far side above the disk mid-plane.
}
\label{posterior_PDF_fig}
\end{Supplementary Figure}

\subsection*{Ionization Equilibrium}

For the material in the disk to be ionized, we need an ionizing photon flux large enough to counteract the recombination loses is needed.
Assuming that the disk is in equilibrium, i.e., that the number of recombinations per unit time is equal to the number of ionizations,
we find
\be
    Q_0 = \int \a_B n_p n_e d^3 r \simeq  \a_B \EM.
\ee
Here $Q_0$ is the flux of ionizing photons with energies $E_\g >13.6$ eV, and
$ \a_B$ is the sum of the recombination coefficients to all levels $\mathrm{n}\geq 2.$ (The so-called Case B recombination: 
recombinations to $\mathrm{n} \geq 2$ result in the 
destruction of an ionizing photon.) Using $\a_B (T=10^4 \textrm{K} ) = 2.59 \times 10^{-13} \, \textrm{cm}^{3},$
we find
\be
    Q_0 =4.6 \times 10^{45}  \left( \frac{\M}{100}\right)^{-1} \,\mathrm{s^{-1}} \label{q0obs}
\ee
or
\be
	L_\mathrm{EUV} = 1 \times 10^{35} \, \mathrm{erg \, s^{-1}} \sim \frac{L_\mathrm{bol}}{30} .
\ee
This is $\sim 1/30$ of the bolometric luminosity of Sgr~A*, which is larger than expected from Sgr~A* \cite{Narayan.Yi.Mahadevan95}. Thus we do not expect all of this ionizing flux to be coming from the black hole itself.

An additional source of ionizing photons is the surrounding stars. We assume that most of the ionizing flux from the stars (S. Ressler and E. Quataert (private communication)) \cite{Martins07,Ressler18} comes from 15 Wolf-Rayet (WR) stars in orbits of $\sim 4$ arcsec. Most of these stars belong to the counterclockwise disk.
The bolometric luminosity of WR stars ranges from $L_\mathrm{WR} \sim 10^{5} - 10^{6} L_{\odot}$ and we expect $\sim 39\% - 69\%$ of this luminosity to be emitted at $E_\gamma \geq 13.6$ eV \cite{Crowther07}.
Near the mean radius of the disk these 15 WR stars produce an ionizing photon flux of
\be
	Q_0^\mathrm{WR} \sim \left(\frac{2}{\pi}\right)^2 \times  \frac{15 \times 10^5 L_\odot \times 0.55}{13.6 \times 1.6 \times 10^{-12}} \times \frac{\pi (0.23 \, \mathrm{arcsec})^2-\pi (0.07 \, \mathrm{arcsec})^2}{4 \pi (4 \, \mathrm{arcsec})^2} \sim 2 \times 10^{46} \,\mathrm{s^{-1}}, \label{q0wr}
\ee
where $(2/\pi)^2$ is a geometrical factor accounting for the disk inclination with respect to the illuminating stars.
Note that $Q_0^\mathrm{WR}$ is an upper limit, since the characteristics of WR stars are uncertain and the emitted EUV would sustain considerable losses before reaching the disk.
Thus we conclude that the ionizing photon flux required to keep the disk ionized is a collective effect of Sgr~A* and nearby orbiting stars.

\subsection*{The disk and the G-objects.}

We now consider the possible effect of our disk of ionized hydrogen on the orbit of the G-objects near Sgr~A*.
The physical properties of G2, and G-objects in general, are uncertain. Some say they are gravitationally unbound gas and dust clouds of a few $M_\mathrm{Earth}$ and $\sim 100$ au radius \cite{Gillessen12,Gillessen18}, while the others argue they are stars embedded in few au dusty envelopes which in turn are embedded in even larger $\Brg$ emitting envelopes and originate from stellar mergers \cite{Witzel14,Witzel17}.
The latter approach avoids the necessity of postulating that G2 assembled itself in its compact form right at the moment when  its detection became possible with introduction of new adaptive optics, it addresses the object's compactness and absence of tidal distortion during the close passage by Sgr~A* in $L'$ broadband filter (which traces the dust emission) \cite{Witzel14,Witzel17}, and naturally explains that its brightness did not change over the decade.

Here we consider a toy model -- a cloud of mass $\sim 3 M_\mathrm{Earth}$ and radius $r_\mathrm{G2} = 0.015 \, \arcsec = 60 \, \mathrm{au}$ \cite{Plewa17,Gillessen18}. In the case that G2 has a cloudy nature this would represent the whole object. In the case of a stellar G2, this would represent its extended $\Brg$ emitting envelope, which takes the hit during the interaction with the disk. The much heavier central star and its surrounding few au dust shell passes through the disk with no noticeable interaction. 
During the collision the disk is treated as stationary. The spherical clouds collide with it at the mean radius of the disk and at a right angle to its plane . 

An encounter with the thin disk (Supplementary Information section ``Disk of a uniform density'') is roughly equivalent to an aluminium ball 1~cm in diameter passing through a 1 mm layer of water, as $r_\mathrm{G2}/H_\mathrm{disk} = 10$ and $n_\mathrm{G2}/n_\mathrm{disk} = 2.9.$
When such a cloud passes through the mean radius of the disk it loses $\sim 1 \%$ of its orbital momentum:
\be
    && \D P_\mathrm{G2} = F_\mathrm{drag} \D t \sim  - \frac{1}{4} \pi r^2_\mathrm{G2} v^2_\mathrm{G2} \frac{2 H}{v_\mathrm{G2}} 
    \sim -P_\mathrm{G2} \times \frac{3}{8} \frac{H}{r_\mathrm{G2}} \frac{\rho_\mathrm{disk}}{\rho_\mathrm{G2}} \sim -\frac{P_\mathrm{G2}}{100},\\
    && \D L_\mathrm{G2} \sim 0.01 L_\mathrm{G2}. 
\ee
Here we used $P_\mathrm{G2} = m_\mathrm{G2} v_\mathrm{G2},$ $L_\mathrm{G2} = R_\mathrm{mean} \times P_\mathrm{G2},$ $F_\mathrm{drag} = -\frac{1}{2} A_\mathrm{eff} C \rho_\mathrm{disk} v^2_\mathrm{G2},$ an effective area of the cloud is $A_\mathrm{eff} = \pi r^2_\mathrm{G2},$ the drag coefficient $C \sim 1/2$ for a ball-like cloud, the duration of the encounter is $\D t = 2 H / v_\mathrm{G2},$ $H/R = 0.01,$  and $R_\mathrm{mean} =0.15 \, \arcsec. $

An encounter with the clumpy BLR-like disk (Supplementary Information section ``BLR-like Ensemble of Clouds'') is roughly equivalent to 1 mm aluminium bullets piercing a 10 cm in diameter water ball, as $r_\mathrm{G2}/r_\mathrm{clouds} \sim 100$ and $n_\mathrm{cloudlets}/n_\mathrm{G2}=2.3$. When G2 passes through the disk composed of bullet-cloudlets it loses $\sim 2\%$ of its orbital momentum:

The loss due to one encounter is
\be
    \D P_\mathrm{G2} = F_\mathrm{drag} \D t \sim  - \frac{1}{4} \pi r^2_\mathrm{clouds} v^2_\mathrm{G2} \frac{r_\mathrm{G2}}{v_\mathrm{G2}} = -P_\mathrm{G2} \times \frac{3}{16} \left( \frac{r_\mathrm{clouds}}{r_\mathrm{G2}} \right)^{2} \sim -P_\mathrm{G2} \times 2 \times 10^{-5}.
\ee
There are $\sim 10^3$ bullet-cloudlets piercing the ball:
\be
    N_\mathrm{collisions} = n_\mathrm{clouds} \times \pi r_\mathrm{G2}^2 \times 2 H \sim \frac{70}{\tan \phi} \times \pi r_\mathrm{G2}^2 \times 2 R_\mathrm{mean} \tan \phi \sim 10^3,
\ee
resulting in a total loss of angular momentum of $\D L^\mathrm{total}_\mathrm{G2} = R_\mathrm{mean} \D P_\mathrm{G2} N_\mathrm{collisions} \sim 0.02 L_\mathrm{G2}.$
This can be less if the internal density of the cloudlets is higher.

The momentum loss of G2 and the disk damage due to the encounter can be zero, if G2 does not interact with the disk at all. This is possible. The closest approach of G2 to Sgr~A* is $\sim 200 \, \mathrm{au} = 0.025 \, \arcsec,$ while the disk has a hole in the middle of the radius size $\sim 0.1 \, \arcsec$ through which G2 can safely pass and avoid the interaction completely. Such a scenario would constrain the disk plane.
The loss of less than a few percent of the G2's orbital momentum due to the encounter with the disk is within the uncertainties of determining G2's orbital parameters \cite{Gillessen18}. 

When the properties of the disk are better determined and the motion of G2 is more accurately constrained, these two objects could constrain properties of each other. At the moment no meaningful constraint can be set on the disk from the motion of G2.

Note that the G1 object faded away $\sim 4-5$ years after its pericenter in 2001, and is believed to have been tidally stripped  \cite{Witzel17}.
The loss of the envelope occurred at a similar distance from Sgr~A* in the plane of the sky as the disk reported here. Unfortunately, no observations prior to or at the time of the G1 close passage are available.


\section*{Author Contributions}

E.M.M. was principal investigator of the observing proposal, analyzed the observational data, conducted theoretical calculations, produced the figures, and wrote most of the paper.
E.S.P. conducted theoretical calculations and made major contributions to the interpretation of  the observational results and to writing the paper.
A.P. modelled the observed spectra, wrote the modelling section, and contributed to interpretation of the observational results.
R.D.B. made a substantial contribution to the interpretation of the observational results.
All co-authors commented on the manuscript.

\section*{Author Information}

Reprints and permissions information is available at www.nature.com/reprints.
The authors declare no competing financial interests. Readers are welcome to comment on the online version of the paper.
Correspondence and request for materials should be addressed to E.M.M. (lena@ias.edu).

\section*{Data availability statement}
This paper makes use of the following ALMA data:
  ADS/JAO.ALMA \#2015.1.00311.S.  \\ 
  The data is publicly available on ALMA archive. 

We use the Common Astronomy Software Applications package (CASA) for the data reduction and analysis.
We use astropy, python and Mathematica for plotting and data analysis.
For the modelling of the spectra, we used the proprietary Code for AGN Reverberation and Modeling of Emission Lines (CARAMEL, \cite{Pancoast14a}). 
The key result presented in this paper is observational. The results of the CARAMEL modelling
are not critical for interpretation of the observational data and
therefore we are not releasing the code with this paper.

\end{document}